\documentclass[aps,pra,amsmath,amssymb,twocolumn,superscriptaddress,showpacs,floatfix]{revtex4-1}
\usepackage{ifpdf}
 \newif\ifpdf
\ifx\pdfoutput\undefined
   \pdffalse
\else
   \pdfoutput=1
   \pdftrue
\fi
\ifpdf
   \usepackage{graphicx}
   \usepackage{epstopdf}
   \usepackage{color}
   \usepackage{verbatim}
\else
   \usepackage{graphicx}
\fi
\usepackage{bm}
\usepackage{srctex}
\usepackage{hyperref}

\DeclareMathOperator{\Real}{Re}
\DeclareMathOperator{\Imag}{Im}

\begin{document}


\title{Comment on ``Influence of image forces on the electron transport in ferroelectric tunnel junctions''}

\author{N.~M.~Chtchelkatchev}
\affiliation{Institute for High Pressure Physics, Russian Academy of Sciences, Moscow (Troitsk) 108840, Russia}
\affiliation{Department of Theoretical Physics, Moscow Institute of Physics and Technology  (State University), Moscow 141700, Russia}
\affiliation{L.D. Landau Institute for Theoretical Physics, Russian Academy of Sciences, Moscow  119334, Russia}
\affiliation{Institute of Metallurgy, Ural Branch, Russian Academy of Sciences, Ekaterinburg 620016, Russia}

\author{A.~V.~Mikheyenkov}
\affiliation{Institute for High Pressure Physics, Russian Academy of Sciences, Moscow (Troitsk) 108840, Russia}
\affiliation{Department of Theoretical Physics, Moscow Institute of Physics and Technology (State University), Moscow 141700, Russia}
\affiliation{National Research Center ``Kurchatov Institute'', Moscow 123182, Russia}

\date{\today}

\begin{abstract}
Udalov and Beloborodov in the recent papers [Phys. Rev. B 95, 134106 (2017); Phys. Rev. B 96, 125425 (2017)] report the strong influence of image forces on the conductance of ferroelectric tunnel junctions. In particular, the authors state that there is enhancement of the electroresistance effect due to polarization hysteresis in symmetric tunnel junctions at nonzero bias. This conjecture seems to be a breakthrough --- the common knowledge is that the considerable effect, linear over voltage bias, takes place only in  NONsymmetric junctions. We show that the influence of image forces on the conductance of ferroelectric tunnel junctions is highly overestimated due to neglecting the difference between characteristic ferroelectric relaxation and electron tunneling times.  We argue that notable enhancement of the electroresistance effect from image forces due to polarization hysteresis in symmetric tunnel junctions at nonzero bias might be observed only at anomalously slow electron tunneling through the barrier. The same applies to magnetic tunnel junctions with a ferroelectric barrier also considered by Udalov et al: there is no significant increase of the magnetoelectric effect due to image forces for typical electron tunneling times. Udalov and Beloborodov completely missed the development of image force theory since 1950's and they forgot that electrons move much faster than atoms in condensed matter. We underline that taking into account dynamical effects in charge tunneling can bring new insight on physics of ferroelectric tunnel junctions.
\end{abstract}


\maketitle
\section{Introduction}
In a recent papers~\cite{Udalov1,Udalov2} Udalov and Beloborodov (UB) address ferroelectric (FE) tunnel junctions where there is ferroelectric layer between metallic electrodes. They investigate nonmagnetic~\cite{Udalov1} and magnetic~\cite{Udalov2} metallic electrodes. In~\cite{Udalov1} UB focused on the special case of symmetric junctions with nonmagnetic equivalent electrodes. Contrary to common knowledge~\cite{Zhuravlev2005PRL,Tsy2010,wen2013ferroelectric,tsymbal2013ferroelectric} UB find the strong enhancement of electroresistance effect taking into account the image force contribution~\cite{burstein1969tunneling,Cole1971PRB,Hanke1985PRB,Viraht2002PRB,Harris1997,gudde2006dynamics} to the tunnel probability. In magnetic tunnel junctions UB show that image forces significantly increase the magnetoelectric effect~\cite{Udalov2}. The predicted effects regarding the symmetric junctions seem very promising not only from academical point of view (the field of research is very relevant~\cite{barrionuevo2014tunneling,tian2016tunnel,li2018resonant,chang2018physical}) but also for advanced microelectronic applications~\cite{tsymbal2013ferroelectric,garcia2014ferroelectric,hou2018ultrathin,dugu2018graphene,hu2018colossal,martinez2018electric,kumari2018ferroelectric,shi2018fast,shi2018fast,huang2018atomically}. So correct understanding is very important.

\begin{figure}[t]
  \centering
  \includegraphics[width=0.9\columnwidth]{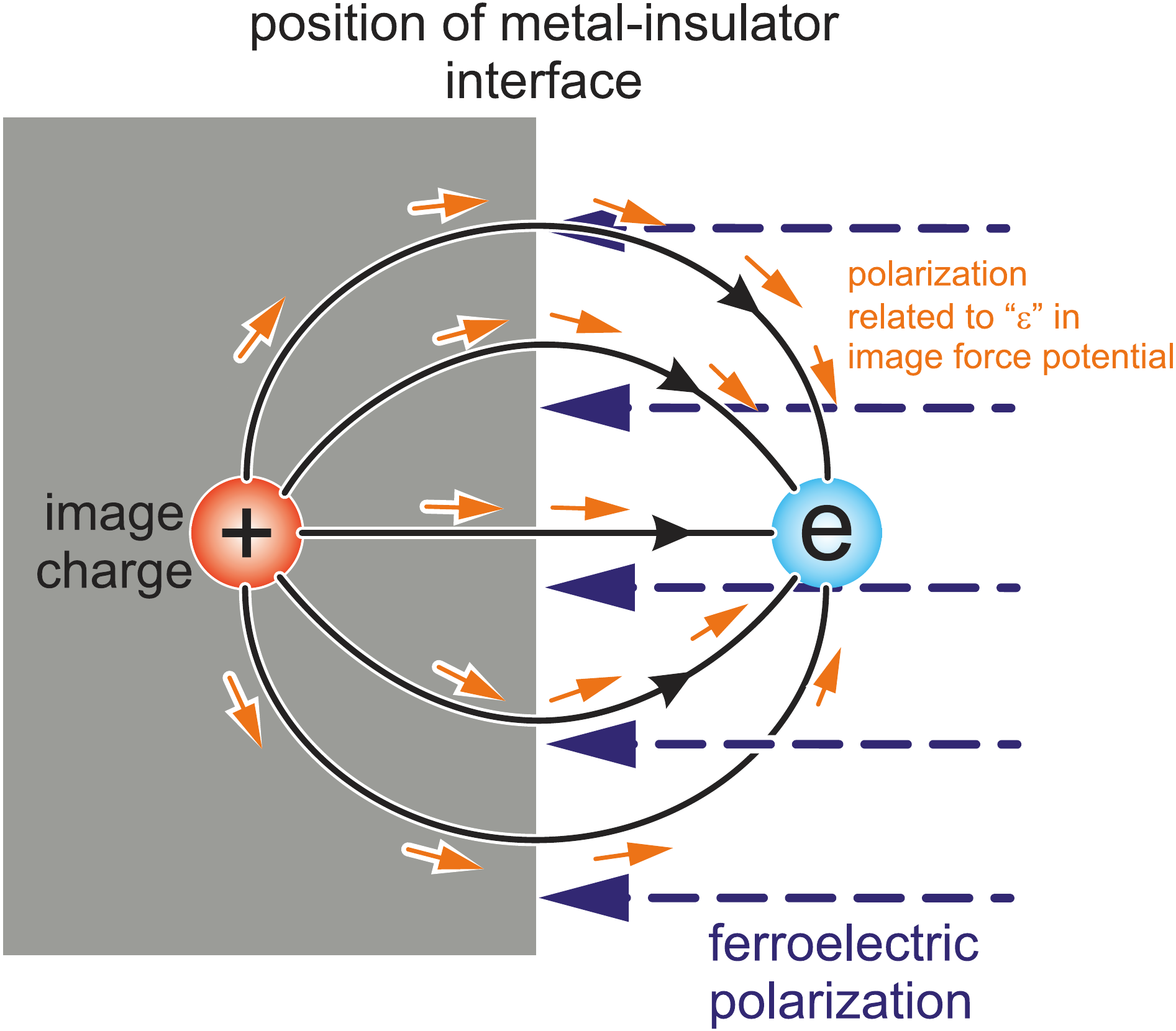}
  \caption{(Color online) Sketch of an electron (``e'') in the insulating barrier of the tunnel junction and one of the image charges (``+'') in the left electrode (the dielectric polarization also has an image on the left side). The right electrode is not shown. Solid black lines with arrows show the distribution of electric field of electron and its image; orange arrows --- polarization of the insulator induced by this electric field. Dashed blue lines correspond to ferroelectric polarization. Note that on the time scales of  electron tunneling the polarization is ``quenched''. }\label{fig1}
\end{figure}

Long ago it has been understood that the potential barrier at the metal-vacuum or metal-insulator interface can not change abruptly as in Gamov model of $\alpha$-decay~\cite{landau2013quantum} because that in fact implies infinite fields~\cite{burstein1969tunneling}. The barrier
really changes smoothly due to the image force. When an electron approaches the surface of a metal from the insulating side it induces the compensating polarization charges that make electric field exactly zero inside the metal. This effect stands behind the origin of  an attractive force (the image force) on the electron~\cite{burstein1969tunneling,Cole1971PRB,Harris1997,Hanke1985PRB,Viraht2002PRB,gudde2006dynamics}:
\begin{gather}\label{eq1}
  F=\frac{e^2}{\epsilon (2x)^2},
\end{gather}
where $x$ is the distance between electron and the metal surface. In Eq.~\eqref{eq1} it is implied that the metal occupies the  half-space while the other half-space is dielectric with  dielectric
constant $\epsilon$, see Fig.~\ref{fig1}. Then the interaction potential $V_0$ due to this image force is equal to
\begin{gather}\label{eq2}
V_0=\int_{\infty}^x F(x)dx=-\frac{e^2}{4 \epsilon x},
\end{gather}
where dielectric occupies the right half-space.

If we have the standard tunnel barrier -- two bulk parallel metallic contacts with dielectric media between them, then infinite series of images appears. The resulting sum for moderate distance $d$ between the metallic contacts is usually approximated by the simple analytical expression~\cite{burstein1969tunneling,Cole1971PRB,Harris1997}:
\begin{gather}\label{eq3}
V\approx-\frac{0.795 e^2 d}{4\epsilon x (d-x)}.
\end{gather}

These considerations are the key point of~\cite{Udalov1,Udalov2}. Most peculiar effects introduced in~\cite{Udalov1,Udalov2} were obtained using Eq.~\eqref{eq3} where $\epsilon$ was related to $\frac{dP_{\rm FE}}{d\mathcal E}$ ($P_{\rm FE}$ is macroscopic polarisation of ferroelectric and $\mathcal E$ is external electric field related to the voltage bias $V$ between the electrodes). This led to the conclusion in~\cite{Udalov1,Udalov2} that $\epsilon=\epsilon(V)$ -- nonlinear function of bias voltage with memory effect mediated by the hysteresis of $P_{\rm FE}(\mathcal E)$.

Below we show that the last statement should be treated very accurately and  effects reported in~\cite{Udalov1,Udalov2} should be revisited and in the presented form they can hardly be observed.

\section{Ferroelectric polarization and electron tunneling}

\subsection{Discussion of the hierarchy of time-scales relevant for a ferroelectric tunnel barrier}

First of all we should note that there is a general fundamental question related to the described style of calculation: how ferroelectric polarization --- macroscopic quantity can enter microscopic calculation like tunneling probability or magnetic exchange interaction. According to modern theory of polarization~\cite{spaldin2012beginner} at microscales $\epsilon$ of a ferroelectric material has pronounces frequency and space dispersion $\epsilon(\omega,k)$~\cite{voitenko2001dynamic}  that is neglected in Refs.~\cite{Udalov1,Udalov2}. Below we put aside this problem and believe that using in any way macroscopic $\epsilon$ in a nanoscale calculation we will extract, like, e.g., in~\cite{Zhuravlev2005PRL,Tsy2010}, physical effects, at least qualitatively. However even then there are problems with approximations done in Refs.~\cite{Udalov1,Udalov2}.

The derivation of Eq.~\eqref{eq3} implies ``adiabatic'' approximation when all the contributions to polarization (related to $\epsilon$) are fast enough~\cite{Thornber1967,buttiker1982traversal,landauer1994barrier,steinberg1995much,WinfulPR2006,KULLIEAP2018} to follow electron moving through the tunnel barrier, see Fig.~\ref{fig1}. In fact polarization consists of several contributions with different characteristic times~\cite{landau2013electrodynamics,feldman2006dielectic}:
\begin{gather}\label{eqP}
\mathbf P=\mathbf P_{\rm el}+\mathbf P_{\rm ion}+\mathbf P_{\rm dipols} + \ldots
\end{gather}
Here the first ``elastic'' contribution is polarization of the outer electron shells, the second one is related to ion shifts, the third is related to dipole moments of molecules etc... It is important that all the contributions except the first one are slow: their relaxation times are larger or of the order of inverse phonon frequencies (with THz, we remind, serving as the natural scale of phonon frequency~\cite{Thornber1967,phonopy,hinuma2017band}). While $P_{\rm el}$ relaxation time  is electronic (optical frequencies) and thus it is much (several orders of magnitude)  shorter. Note, also, that the ``slow'' terms in~\eqref{eqP} produce the leading contribution to $P_{\rm FE}$.
\begin{figure}[tb]
  \centering
  \includegraphics[width=\columnwidth]{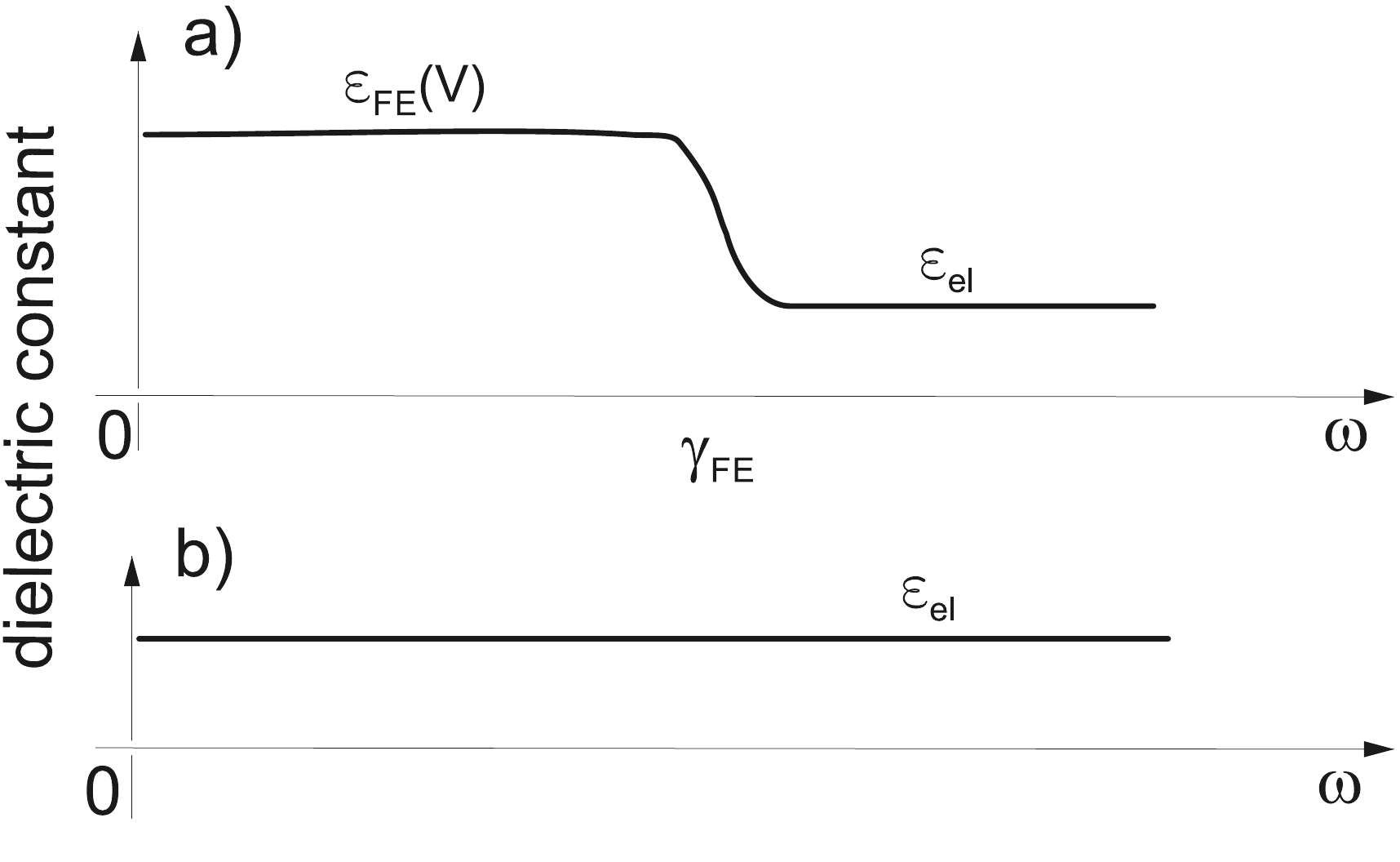}
  \caption{(Color online) Figure (a) illustrates the typical frequency behaviour of ferroelectric differential dielectric constant with frequency: ionic, dipole etc... contributions die at large frequencies  and only electronic, the so called elastic contribution, survives. Figure (b) illustrates that purely electronic contribution nearly does not depend on frequency (up to optical range). Here only the real part of  dielectric constant is considered and all  $\epsilon(\omega)$ ``fine structure'' is omitted so some envelop function is shown.
  \\
  Note that ionic, dipole etc... contributions depend on constant (zero frequency component) voltage bias, $V$, while the electronic contribution does not.}\label{fig2}
\end{figure}

Relaxation dynamics of the ferroelectric order parameter can be estimated from
\begin{gather}\label{eqPt}
\gamma_{\rm FE} \mathbf{{\dot P_{\rm FE}}}=-\frac{\delta F_{\rm LD}[\mathbf P_{\rm FE}]}{\delta \mathbf P_{\rm FE}}+\mathcal E_{\rm ext}(t),
\end{gather}
where $\gamma_{\rm FE}=1/\tau_{\rm FE}$ is the inverse relaxation time of ferroelectric polarization (order parameter), $F_{\rm LD}[\mathbf P_{\rm FE}]$ is the Landau-Devonshire free energy~\cite{landau2013electrodynamics} that describes ferroelectric, and $\mathcal E_{\rm ext}(t)$ is time-dependent external electric field.

If we take $\mathcal E_{\rm ext}(t) \sim \mathcal E_{\rm ext}^{(0)}  e^{-i\omega t}$ with $\omega$ much larger than any characteristic frequency of a ferroelectric, then $F_{\rm LD}$-term becomes irrelevant in Eq.~\eqref{eqPt},  in the Fourier space $P_\omega \sim \mathcal E_{\rm ext}^{(0)}/(-i\omega \gamma_{\rm FE})$, and, thus, $\epsilon(\omega) \sim 1/(-i\omega \gamma_{\rm FE})$. This is very rough estimate that only illustrates the well known behaviour of ferroelectric dielectric constant with frequency: ferroelectricity does not respond on large enough frequencies. This is sketched in Fig.~\ref{fig2}.



\subsection{Discussion of image forces and the conductance of ferroelectric tunnel junctions }

We can conclude following Refs.~\cite{Thornber1967,Heinrichs1973PRB,Baratoff1988PRB} about $\epsilon$ in Eq.~\eqref{eq3} -- the key equation of Refs.~\cite{Udalov1,Udalov2} that only the elastic (electron) contribution ``works'':
\begin{gather}\label{eqeps}
\epsilon\approx\Real \epsilon(\omega=\omega_{\rm tn})\approx\Real \epsilon_{\rm el}(\omega=\omega_{\rm tn})\approx  \epsilon_{\rm el}(\omega=0),
\end{gather}
where $\omega_{\rm tn}=1/\tau_{\rm tn}$ and $\tau_{\rm tn}$ is the time scale of the order of electron tunneling time~\cite{Thornber1967,buttiker1982traversal,landauer1994barrier,steinberg1995much,WinfulPR2006,KULLIEAP2018}.
This $\epsilon$ is not known to be notably depending  on ferroelectric polarization in the tunnel junction (and voltage bias as well unless the voltage produces the fields of the order of intrinsic atomic fields) and as the consequence, the effects predicted in Refs.~\cite{Udalov1,Udalov2} are under question and require revision.

To be more specific, we examine below the key equations of Refs.~\cite{Udalov1} for image-force contribution to the resistance of the ferroelectric tunnel junction. UB assume that the FE barrier is thin enough and the electron transport occurs due to tunnelling. UB calculate electric current $I$ across the barrier using the Simmon's formula derived in 1963~[see~\cite{Simmons1963,Simmons1963_1}] for tunnel junctions with large area electrodes [quasiclassical tunneling formula where the integral over transverse momentum has been explicitly carried out]:
\begin{equation}\label{Eq:Curr}
I=J_0(\overline U(h_\mathrm b) e^{-A\sqrt{\overline U(h_\mathrm b)}}-\overline U(h_\mathrm b+V)e^{-A\sqrt{\overline U(h_\mathrm b+V)}} ),
\end{equation}
where $\sqrt{\overline U}=\frac{1}{d_\mathrm{eff}}\int_{z_1}^{z_2}\sqrt{U(z)}dz$, and $U(z)$ is the profile of the effective potential barrier of the tunnel junction, $z_{1,2}$ are the ``turning points'' where $U(z)=0$,
the parameter $A=\beta d_\mathrm{eff}\sqrt{2m_\mathrm e e/\hbar^2}$ and $J_0=(e^2/\hbar\beta d_\mathrm{eff}^2)$. Here $\beta\sim 1$, and $d_\mathrm{eff}=z_2-z_1$ is the effective thickness of the barrier~\cite{Udalov1}.

One of the terms in $U(z)$ taken into account by UB in Refs.~\cite{Udalov1} is the image force potential~\eqref{eq3}. UB show that the contribution of the image forces into the average of $U(z)$ is
\begin{equation}\label{eq123}
d_\mathrm{eff}\sqrt{\overline{U}}\approx d\sqrt{h_\mathrm b}\left(1-\frac{h_\mathrm c}{2h_\mathrm b}\left(1+\frac{1}{2}\mathrm{ln}\frac{h_\mathrm b}{4h_\mathrm c}\right)\right),
\end{equation}
where $h_b$ defines in~\cite{Udalov1}  the barrier height above the Fermi level $E_F$ of the left lead ($E_F=0$  in~\cite{Udalov1}) in the absence of FE polarization, image forces and external voltage. Important parameter $h_c=0.795 e/\epsilon d$  is the characteristic potential associated with image forces.

UB connect $\epsilon$ in $h_c$ with the differential polarizability of the ferroelectric and using Eq.~\eqref{eq123} arrive at the conclusion that the image force contribution to $d_\mathrm{eff}\sqrt{\overline{U}}$ depends on the polarization orientation of the ferroelectric in the tunnel barrier and, more important, follows its hysteresis:
\begin{equation}\label{Eq:EstImF}
d_\mathrm{eff}\sqrt{\overline{U}}|_{P^+}-d_\mathrm{eff}\sqrt{\overline{U}}|_{P^-}\sim d\sqrt{h_\mathrm b} \frac{h^0_\mathrm c}{h_\mathrm b}\frac{\epsilon^+-\epsilon^-}{\epsilon^+ \epsilon^-}\neq 0,
\end{equation}
where $\pm$ label the upper (lower) hysteresis branch of FE according to Ref.~\cite{Udalov1}.

$\epsilon^\pm$  in Eq.~\eqref{Eq:EstImF} originate from $\epsilon$ in $h_c$ according to~\cite{Udalov1}. However we have argued above, see Eq.~\eqref{eqeps}, that $\epsilon$ can hardly be sensitive to hysteresis branch of FE layer since is represents only elastic (electron) contribution to the dielectric constant at high frequencies, much higher than any inverse relaxation time of FE. So we conclude that contrary to~\eqref{Eq:EstImF} stated in~\cite{Udalov1}, according to Eq.~\eqref{eqeps}, $\epsilon^+=\epsilon^-\approx \epsilon_{\rm el}$ and the truth is
\begin{equation}\label{Eq:EstImFcorrect}
\left(d_\mathrm{eff}\sqrt{\overline{U}}|_{P^+}-d_\mathrm{eff}\sqrt{\overline{U}}|_{P^-}\right)_{\rm image\, force}\approx 0.
\end{equation}

It already follows from Eq.~\eqref{Eq:EstImFcorrect} and arguments given above that all effects related to the interplay of FE hysterisis and image forces in~\cite{Udalov1} are overestimated. In particular, image forces do not give  significant contribution to electroresistance effect (ER).

\begin{figure}[tb]
  \centering
  \includegraphics[width=\columnwidth]{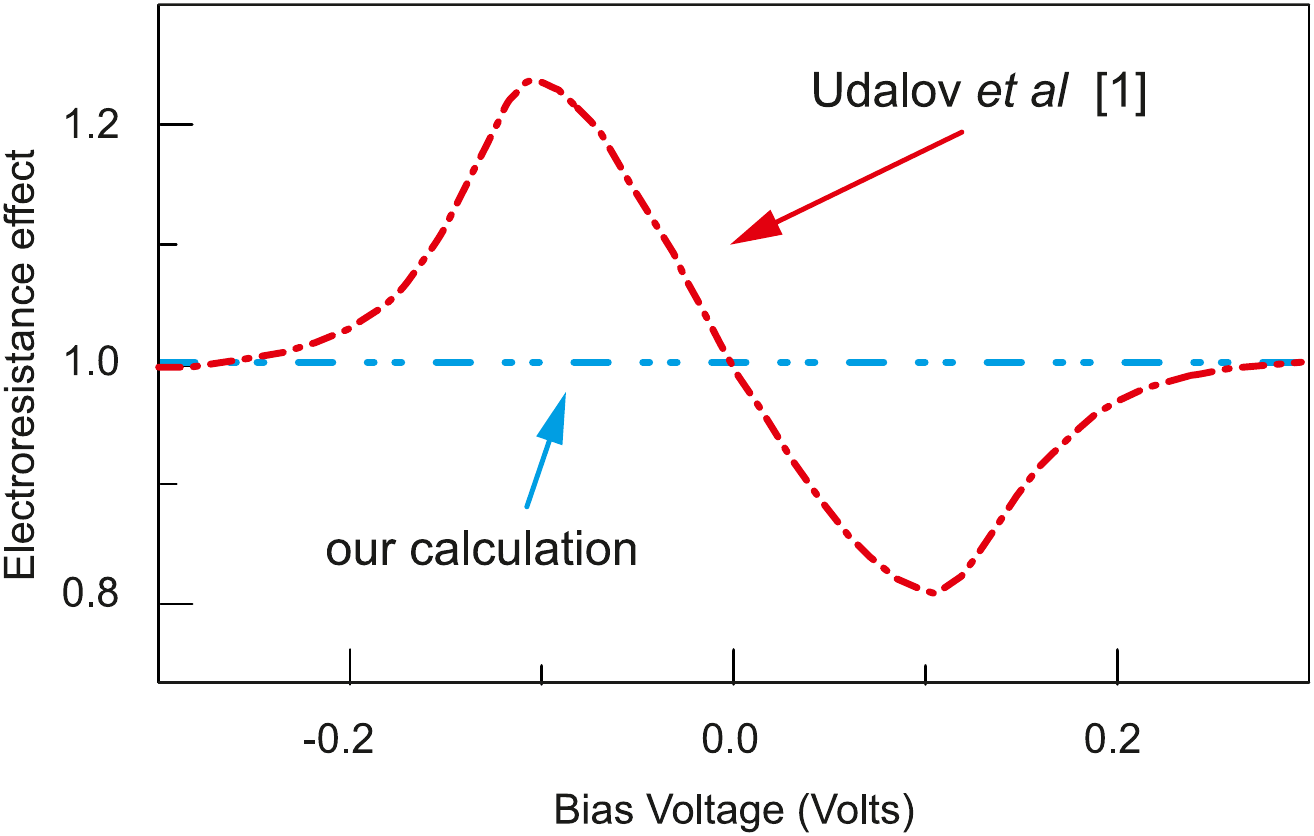}
  \caption{(Color online)  Electroresistance effect (ER) due to image forces  as a function of applied voltage for the following system parameters like in Fig.3 of Ref.~\cite{Udalov1}: $d = 1$~nm, $h_b = 0.5$~eV, $P_0 = 30$~$\mu$C/cm$^2$, $\delta_1 = \delta_2 = 0.05$~nm, $V_s = 0.1$~V, $\Delta ε = 15$, and $\epsilon_{\rm min} = 30$. Our calculation shows no detectable electroresistance effect mediated by image forces (blue dash--dot--dot line) while UB observe the effect (red dash--dot line, see Fig.3 in Ref.~\cite{Udalov1}) because UB relate $\epsilon$ in~\eqref{eq2} with  $\frac{dP_{\rm FE}}{d\mathcal E}$ and conclude that $\epsilon=\epsilon(V)$ is the nonlinear function of bias voltage with memory effect mediated by the hysteresis of $P_{\rm FE}(\mathcal E)$. Our message that  $\epsilon$ in~\eqref{eq2} is  related only  the fast ``electron'' contribution to polarization and therefore it does not show the pronounced hysteresis and so we see no detectable ER effect.}\label{fig3udalov}
\end{figure}

However, for clarity,  in addition to checking asymptotic approximations made in~\cite{Udalov1}, we also recalculate numerically observables that UB~\cite{Udalov1} represent as key results.  For example, we calculated numerically electroresistance effect (ER) due to image forces as a function of applied voltage using Eq.~\eqref{Eq:Curr} like in~\cite{Udalov1}, but taking into account that $\epsilon$ in the image force potential is actually some constant ($\epsilon_{\rm el}$) of the order of unity that does not depend on  FE hysterisis. Our results are shown in Fig.~\eqref{fig3udalov} where they are compared with the results of~\cite{Udalov1}: we obviously do not notice a contribution to ER due to image forces.

\subsection{Discussion of image forces influence on the interlayer exchange interaction in magnetic tunnel junctions with a ferroelectric barrier}

In Ref.~\cite{Udalov2} UB study the interlayer exchange interaction in magnetic tunnel junctions with a ferroelectric barrier focusing on the influence of image forces on the voltage dependence of the interlayer magnetic interaction (magnetoelectric effect). In the beginning of~\cite{Udalov2} UB write Eq.~\eqref{eq3} for the image force as the starting point and again as in Ref.~\cite{Udalov1} UB do directly connect $\epsilon$ in the image force potential to $\frac{dP_{\rm FE}}{d\mathcal E}$. Then UB write the expression for $h_c$ (given here above) and explain that $h_c$ strongly differs depending on FE hysteresis branch. From that point UB report about  a number of peculiar effects related to the interplay of FE hysteresis and interlayer exchange interaction mediated by image forces.

We have already explained above why $\epsilon$ in the image force potential is actually $\epsilon_{\rm el}$ and why it can not be simply related to $\frac{dP_{\rm FE}}{d\mathcal E}$. From that point we can conclude without a shadow of doubt that most results reported in~\cite{Udalov2}, related to FE hysteresis and image forces (the key results of~\cite{Udalov2}), are strongly overestimated like it was in~\cite{Udalov1}.

\section{Importance of dynamical effects in charge tunneling through active dielectric layers (including ferroelectric)}

Ideas developed in Refs.~\cite{Udalov1,Udalov2} are interesting but due to overestimates require a revision.  There are several ways of such a revision: one way corresponds to significant increase of the frequency response range of a ferroelectric and the other --- to significant slowdown of electron tunneling~\cite{Thornber1967}. It is known that after tunneling, some time is required for the diffusion of extra electric charge over the electrode~\cite{levitov1997semiclassical}, may be this will help. However  all these opportunities are challenging for an experiment.

But there is also another option. UB investigate the conductance and believe that it is proportional to, roughly speaking, the square absolute value of the tunnel amplitude, $|t|^2$. This description of the conductance is not accurate enough. The amplitude $t$ has also phase $\phi$, $t=|t|e^{i\phi}$. It is an important parameter.

If we oversimplify the physical picture, FE is the sequence of nonlinear oscillators that oscillate due to temperature (coupling to phonon bath) and electron current going through the tunnel barrier. Tunneling electron may exchange energy with the ``bath'' (or ``environment'') of these oscillators,  giving or receiving energy. Physically, this is slightly similar to inelastic tunneling phenomena, e.g., phonon mediated, investigated long ago~\cite{burstein1969tunneling}. The phase of the tunnel amplitude is sensitive to the environment quantum state. So calculation of the conductance in the tunnel junction with active dielectric inside should take into account the fluctuations of the tunneling phase $\phi$ that couples tunneling electron with active dielectric.

Let us return to the tunnel junction  with a dielectric inside, but having $\epsilon$ with a frequency dependence (this might be a ferroelectric as well). Then the capacitance of the junction is also frequency dependent, $C=C(\omega)$. If we apply the voltage bias $V$ to this junction then its current-voltage characteristics can be found in a standard way:
\begin{gather}\label{eqIV0}
I(V)=\overrightarrow{\Gamma}(V)- \overleftarrow{\Gamma}(V),
\end{gather}
where $\overrightarrow{\Gamma}(V)$  and $\overleftarrow{\Gamma}(V)$ are electron forward and backward tunnel rates  (we take mostly everywhere below units where $e=1$, $\hbar=1$ and $k_B=1$).

For simplicity here we focus on the tunnel junctions with perfect metallic leads and consider small enough bias voltage. The densities of states in the leads can be approximately taken at the Fermi levels and the reshape of the tunnel barrier due to electric field can be neglected. Then~\cite{DevoretPRL1990,devoret1992single}
\begin{multline}\label{eqGamma}
\overrightarrow{\Gamma}(V)=
\\
\frac{1} { R_T} \int_{-\infty}^\infty dEdE' f_1(E)[1 - f_2(E' + V)]P(E - E') =
\\
\frac{1} { R_T} \int_{-\infty}^\infty d\omega N_{12}(\omega) P(\omega),
\end{multline}
where $N_{12}(\omega)=\int dE f_1(E+\omega)[1-f_2(E)]=[\omega-V]N_B[\omega-V]$, $N_B(\omega)$ is the Bose function.
Expression for $\overleftarrow{\Gamma}(V)$ has permuted indices $1,2$ compared to Eq.~\eqref{eqGamma}.
Here  $R_T$ is the bare tunneling resistance ($\propto |t|^{-2}$) , $f_{1,2}$ are the electron distribution functions in the left (right) electrode, $E$ is electron energy. Here we believe that the electrodes are in local equilibrium, so $f_{1,2}(E)=\frac12[1-\tanh(E/2T)]$ is the Fermi function where $T$ is temperature.

$P(E - E')$ is the probability that tunneling electron shares the energy $E - E'$ with the  ``environment'' during the tunneling process. If there is no environment, tunneling is elastic then $P(E - E')=\delta(E - E')$ as in the Fermi golden rule and Eq.\eqref{eqGamma} becomes the linear in $V$ approximation of~\eqref{Eq:Curr}. The most important thing is that this probability $P$ is built from the time correlation functions of the tunnel amplitude phases $\phi(t)$ that we discussed above~\cite{DevoretPRL1990,devoret1992single}.

We can rewrite Eq.~\eqref{eqGamma} in the time representation using the Fourier transform:
\begin{gather}\label{eqPp}
P(E)= \frac{1}{2\pi}\int_{-\infty}^\infty dt \exp( J(t) +i Et),
\end{gather}
and so $\overrightarrow{\Gamma}=\frac{1}{R_T}\int_{-\infty}^\infty dt N_{12}(-t) e^{J(t)}$.
The Fourier transform of $N_{12}(\omega)$:
\begin{gather}
 N_{12}(t)=-\frac1{2\pi}\frac{(\pi T)^2}{\sinh^2(\pi T t)}e^{-iV t}-\frac{i}2 \delta'(t)+\frac 12 eV\delta(t).
\end{gather}
Then we finally arrive at
\begin{gather}\label{eqIt}
 I=\frac{ V}{R_T}+\frac{2}{\pi  R_T}\int_{0}^\infty dt \frac{(\pi T)^2}{\sinh^2(\pi T t)} \sin(V t)\,\Imag e^{J(t)}.
\end{gather}

Eq.~\eqref{eqIt} is not always very convenient for practical model calculations. Sometimes more convenient is to work in $\omega$-representation~\eqref{eqGamma}.  But this expression has quite transparent physical interpretation if we search for relevant time-scales. So, in the absence of any environment, tunneling is nearly instant process and we have the conventional Ohm's law, $ I(V)= V/R_T$, as follows from~\eqref{eqIt}. The environment makes $I(V)$ nonlinear, also it brings in a number of time-scales.  Clearly there is a competition between $\hbar/k_B T$, $\hbar/eV$ and time scales of $\Imag e^{J(t)}$ related, from one hand, to dielectric constant characteristic time-scales and, from the other hand, to some dynamical effects of the charge transfer process.

Strictly speaking Eqs.~\eqref{eqPp}-\eqref{eqIt} can be considered as some semiquantitative example showing the influence of polarization frequency dispersion on the charge transport through the tunnel junction. If we take more general case, the final result would be like  Eq.~\eqref{eqIt}: the Ohm's law (or some nonlinear generalisation like Eq.~\eqref{Eq:Curr} mediated by ``static'' physics) plus some nontrivial nonlinear contribution induced by time-dependent phenomena.

We should note that \eqref{eqIt} formally  is not restricted to ultrasmall junctions. However for ultrasmall  tunnel junctions $P(E)$ can be rather easily expressed through the total impedance $ Z_t(\omega)$ of electromagnetic environment as follows~\cite{DevoretPRL1990,devoret1992single}:
\begin{multline}\label{eqJ}
J(t) =2 \int_0^\infty \frac{d\omega}\omega\frac{\Real Z_t(\omega)}{R_Q}\times
\\
\left\{ \coth ( \omega/2T ) [\cos(\omega t) - 1] - i \sin(\omega t)\right\},
\end{multline}
where $R_Q$ is the resistance quantum  and
\begin{gather}\label{eqZ}
 Z_t(\omega) =\frac1{ i\omega C(\omega) + Y_{\rm env}(\omega)}.
\end{gather}
Here $Y_{\rm env}$ is the impedance shunting the tunnel junction. In our case,  $Y_{\rm env}(\omega)=0$ (or $|Y_{\rm env}(\omega)|\ll |\omega C(\omega)|$), so all the environment is represented by the frequency dependent capacitance of the tunnel junction. This expression for $J(t)$ is valid while $R_Q\ll R_T$~\cite{DevoretPRL1990,devoret1992single}.

Returning to \eqref{eqIt} late us take, as the toy model, the simplest Drude-model~\cite{feldman2006dielectic,ye2008handbook,poplavko2009physics} for $\epsilon(\omega)$ dependence, that corresponds to Fig.~\ref{fig2} (below we identify $\epsilon_{\rm el}$ with $\epsilon_\infty$):
\begin{gather}\label{eqC}
C(\omega)=C_0\left\{\epsilon_\infty+\frac{\epsilon_0-\epsilon_\infty}{1+i\omega\tau}\right\},
\end{gather}
where $\tau$ is the relaxation time of $\epsilon(\omega)$ and $C_0$ is the ``geometrical'' capacitance $\sim S/d$ ($S$ is of the order of the junction area and $d$ is the characteristic distance between the electrodes). Then
\begin{gather}\label{eqReZt}
  \Real Z_t(\omega) =\frac{-\Imag \epsilon(\omega)}{\omega C_0|\epsilon(\omega)|^2}=\frac{\tau  \left(\epsilon_0-\epsilon_{\infty }\right)} {C_0 \left(\epsilon_0^2+(\omega \tau\epsilon_{\infty })^2\right)},
\end{gather}
and finally
\begin{gather}
\frac{  \Real Z_t(\omega)}{R_Q}=\frac{1}g\frac{1} { 1+(\omega /\omega_R)^2},
\\\label{eqomegaR}
g=\frac{\epsilon_0^2 C_0 R_Q}{\tau \left(\epsilon_0-\epsilon_{\infty }\right)},\qquad \omega_R=\frac{\epsilon_0}{\epsilon_\infty}\frac1\tau.
\end{gather}
This notations map the problem in hand to the well known case of, so-called, ``Ohmic'' environment.

Then for zero temperature, $|V|<\omega_R$, the current-voltage characteristic is essentially nonlinear~\cite{Girvin1990PhysRevLett,DevoretPRL1990,devoret1992single}:
\begin{gather}\label{eqIV}
  I(V)\approx \frac V{R_T}\left\{\frac{|V|}{\omega_R}\right\}^{2/g}\frac{\exp( -2\gamma/ g)}{\Gamma(2+2/g)}.
\end{gather}
Thus, frequency dependent capacitance leads to a zero-bias anomaly of the conductance $dI/dV\sim V^{2/g}$.

The long time asymptotic of $J(t)$ is mostly responsible for Eq.~\eqref{eqIV}:
\begin{gather}\label{eqJt}
J(t)\approx -\frac 2 g\left[ \ln (\omega_R t)+i\frac\pi 2+\gamma\right],
\end{gather}
where $\gamma=0.577...$ is the Euler constant. However direct derivation of~\eqref{eqIV} in the time representation is a bit tricky  because all time scales will be actually involved, not only the long time scales. One should check that the first Ohmic term in~\eqref{eqIt} exactly cancels with the certain part coming from the second term in~\eqref{eqIt} and only after that one arrives at~\eqref{eqIV} using something like ~\eqref{eqJt}.  The derivation of~\eqref{eqIV} is much easier in the $\omega$-representation where at zero temperature, $R_T I(V)= \int_0^V(V-\omega)P(\omega)d\omega$~\cite{Girvin1990PhysRevLett,DevoretPRL1990,devoret1992single} [it follows from Eqs.~\eqref{eqIV0},\eqref{eqGamma} and the condition: $P(\omega<0)=0$ if $T=0$] and $P(\omega)\sim \omega^{2/g-1}$.

For $V\gg \omega_R/g$, the short time asymptotic is the most important
\begin{gather}\label{eqJt2}
\Imag P(t)=\Imag e^{J(t)}\approx-\frac{\pi\omega_R}{g} t+\frac{\pi \omega_R^2}{2g^2}t^2,
\end{gather}
that with the help of  Eq.~\eqref{eqIt} finally produces the second and the third terms below:
\begin{gather}\label{eqIV11}
  I(V)\approx \frac 1{R_T}\left\{V-\frac{\pi\omega_R}g+\frac{\omega_R^2}{g^2V}\right\}.
\end{gather}

The time scales $\omega_R^{-1}$ and $g/\omega_R$  defined in~\eqref{eqomegaR},\eqref{eqJt} and \eqref{eqJt2} are characteristic times relevant for charge transport through the tunnel junction. However they should not be confused with the time of tunneling through the tunnel barrier. We should remind that the charge transport in the tunnel junction goes roughly speaking in several stages, where the first one is very quick quantum-mechanical tunneling  through the tunnel barrier and the second one is related to relatively slow fluctuation of electric field generated a) by an electron-hole pair: electron in the ``drain'' electrode and hole in the source electrode, and b)  by excitation of active dielectric. These processes somehow are built in the second term of Eq.~\eqref{eqIt}.

Taking $C_0=0.1$~fF like in \cite{Girvin1990PhysRevLett} and $1/\tau=1$~THz, we get $g\sim 2.5$  and $\omega_R\sim 0.04\,\mathrm{eV}=48\,\mathrm K$. (Today tunnel junctions with $C_0\leq0.01$~fF can be experimentally prepared that leads to $g\leq0.25$ and $\omega_R\approx 480$~K.) A bit tricky is find material with $1/\tau=1$~THz -- it is the upper boundary to $1/\tau$. However even if $1/\tau=10-100$~GHz  than effects discussed here might be observable.

In Eqs.~\eqref{eqJ},\eqref{eqIV}-\eqref{eqIV11} we focused on ultrasmall tunnel junctions. Small values of $C_0$ were required above only to ensure reasonably large $\omega_R$. Generalisation to tunnel junctions with arbitrary large area could be made if we consider the tunnel junction as the circuit with an array of  parallel ultrasmall tunnel junctions. This consideration we leave for the forthcoming paper.

More interesting and relevant than the Drude model~\eqref{eqC} is the Lorentz (or ``oscillator'') feature in the dielectric function spectrum~\cite{volkov2003broadband,feldman2006dielectic,ye2008handbook,poplavko2009physics}:
\begin{gather}\label{eqepsL}
  \epsilon(\omega)=\epsilon_\infty+\frac{\epsilon_0-\epsilon_\infty}{1-(\omega/\omega_0)^2-i(\omega/\omega_0) \Gamma },
\end{gather}
where $\omega_0$ is oscillator frequency (e.g., ferroelectric resonance frequency) and $ \Gamma$ is the ratio of damping and $\omega_0$. It intuitively clear that something interesting will happen around $V\sim\omega_0$ in $I(V)$ characteristic. We also leave this investigation for forthcoming publications.

If we deal with ferroelectric layer in the tunnel junctions then  $\epsilon_0=\epsilon(\omega=0)$ (the static dielectric constant) in \eqref{eqC} and \eqref{eqepsL} could be related with  $\frac{dP_{\rm FE}}{d\mathcal E}$. Thus $\epsilon_0$ depends on the hysteresis branch and we will see some electroresistance effect in $I(V)$.

\section{Discussion and Conclusions}

We have shown, following the dynamical theory of image force effect~\cite{Thornber1967,Heinrichs1973PRB,Baratoff1988PRB}, that correct description of kinetics in tunnel junctions with active dielectric (or ferroelectric) layers requires understanding hierarchy of time-scales related to the dynamics of charge transfer and dynamics (relaxation) of polarization.

We have found that  there is no noticeable influence of image forces on electroresistance and magnetoelectric effect in ferroelectric tunnel junctions contrary to investigations in Refs.~\cite{Udalov1,Udalov2}.

Udalov and Beloborodov missed that since the publication of the book "Tunneling phenomena in solids"~\cite{burstein1969tunneling} 60 years ago, tunneling physics including image force theory has advanced a lot~\cite{Heinrichs1973PRB,Baratoff1988PRB}. Most important, they missed that electrons move so fast in condensed matter that one tunneling electron can hardly make an atom of the insulating layer shift during the single tunneling event~\cite{landau2013quantum}.

The mentioned problems are not limited to just two ``papers''~\cite{Udalov1,Udalov2} of Udalov and Beloborodov: in fact, most papers of these team published last time about the so-called ``granular multifferroics'' and magnetoresistance effect  are the same.

\begin{acknowledgments}
This work was supported by the program 0033-2018-0001 ``Condensed Matter Physics'' by the FASO of Russia and partly by the Russian Foundation for Basic Research (projects No. 16-02-00295).
\end{acknowledgments}
\bibliography{mybibfile}
\end{document}